\documentclass{appolb}
\usepackage{epsfig}
\usepackage{wrapfig}
\newcommand{\lam}{\mbox{$\rm \Lambda ^0$}}
\newcommand{\alam}{\mbox{$\rm \bar \Lambda ^0$}} 

\begin{document}
\title{Intrinsic Polarized Strangeness and $\lam$ Polarization in 
Deep-Inelastic Production%
\thanks{This article is a short version of a more detailed paper~\cite{EKN}.
}%
}
\author{Dmitry V.~Naumov
\address{JINR, Dubna, Russia}
}
\maketitle
\vspace*{-1cm}
\begin{abstract}
We propose a model for the longitudinal polarization of $\lam$ baryons
produced in deep-inelastic lepton scattering at any $x_F$, based on
static $SU(6)$ quark-diquark wave functions and polarized intrinsic
strangeness in the nucleon associated with individual valence quarks. Free
parameters of the model are fixed by fitting the NOMAD data on the
longitudinal polarization of $\lam$ hyperons in neutrino interactions. Our
model correctly reproduces the observed dependences of $\lam$ polarization
on the kinematic variables. Within the context of our model, the NOMAD
data imply that the intrinsic strangeness associated with a valence
quark has anticorrelated polarization. We also compare our model
predictions with results from the HERMES and E665 experiments using 
charged leptons. Predictions of our model for the COMPASS experiment are 
also presented.
\end{abstract}
\section{Introduction}
Measurements of the longitudinal polarization of $\lam$ hyperons in
lepton nucleon deep inelastic scatering (DIS) processes provide access
to the polarization of intrinsic strangeness of the nucleon~\cite{polstr} 
and to the polarized quark spin transfer function~\cite{BJ}:
$C_q^\Lambda(z) \equiv \Delta D_q^\Lambda(z)/D_q^\Lambda(z)$,
where $D_q^\Lambda(z)$ and $\Delta D_q^\Lambda(z)$ are unpolarized and
polarized fragmentation functions for the quark $q$ to yield a $\Lambda$
hyperon with the fraction $z$ of the quark energy.
Several experimental measurements of $\lam$ polarization have been made in
neutrino and anti-neutrino DIS. Longitudinal polarization of $\lam$
hyperons was first observed in bubble chamber (anti) neutrino
experiments~\cite{wa21,wa59,e632}. 
The NOMAD Collaboration has recently published new and interesting
results on $\lam$ and $\alam$ polarization with much larger
statistics~\cite{NOMAD_lambda_polar}. 
There are also recent results on longitudinal polarization of $\lam$ hyperons
from polarized charged lepton nucleon DIS processes  from the
E665~\cite{e665} and HERMES~\cite{hermes} experiments.
A key assumption, adopted widely in theoretical analyses of these data, 
is that the struck quark fragmentation can be disentangled from the 
nucleon remnant fragmentation by imposing a cut: $x_F>0$. As we show in 
Sec.~\ref{sec:frag_model} this assumption fails at moderate beam 
energies~\cite{wa21,wa59,e632,NOMAD_lambda_polar,e665,hermes,compass}. 
A method of calculation of the longitudinal polarization of $\lam$
hyperons is presented in Sec.~\ref{sec:method}, and our model predictions are 
compared to the available data in Sec.~\ref{sec:results}.\\[-1cm]

\section{Calculational Method\label{sec:method}}
There are different mechanisms whereby strange hadrons can be produced
in DIS processes. They can be produced by fragmentation of the struck quark or
the nucleon remnant diquark, or in color string fragmentation.
We assume that strange hadrons can be polarized only in the (di) quark
fragmentation. $\lam$ hyperons can be produced {\em promptly} or as a 
decay product of heavier strange baryons 
($\Sigma^0$, $\Xi$, ${\Sigma}^\star$).
Therefore,  to predict the polarization of $\lam$ hyperons in a 
given kinematic 
domain one needs to know the relative yields of $\lam$'s produced in 
different channels and their polarization. We take into account all these
effects explicitly tracing the $\lam$ origin predicted by the fragmentation 
model adopted (Sec.~\ref{sec:frag_model}) and assigning the polarization 
predicted by the polarized intrinsic strangeness model (Sec.~\ref{sec:model})
in the diquark fragmentation and by SU(6) and Burhardt-Jaffe~\cite{BJ} 
models for the quark fragmentation (Sec.~\ref{sec:lam_polar}). We take into 
account a difference in observed yields of heavier strange 
hyperons~\cite{Astier:2001vi}.
\\[-1cm]

\subsection{Polarized Intrinsic Strangeness Model\label{sec:model}}
The main idea of the polarized intrinsic strangeness model applied to
semi-inclusive DIS is that the polarization of $s$ quarks and $\bar s$ 
antiquarks in the hidden strangeness component of the nucleon wave 
function should be (anti)correlated with that of the struck quark. This 
correlation is described by the spin correlation coefficients $C_{sq}$:
$P_s=C_{sq} P_q$,
where $P_q$ and $P_s$ are the polarizations of the initial struck
(anti)quark and remnant $s$ quark. In principle, $C_{sq}$ can be different for
the valence and sea quarks. We leave $C_{sq_{val}}$ and $C_{sq_{sea}}$ as free
parameters, that are fixed in a fit to the NOMAD 
data~\cite{NOMAD_lambda_polar}.\\[-1cm] 

\subsection{Polarization of Strange Hadrons in (Di)quark 
fragmentation\label{sec:lam_polar}}
We define the quantization axis along the three-momentum vector of the
exchanged boson. 
To calculate the polarization of ${\lam}$ hyperons produced in the 
diquark fragmentation we assume the combination
of a non-relativistic $SU(6)$ quark-diquark wave function and the
polarized intrinsic strangeness model described above.
The polarization of ${\lam}$ hyperons produced in the 
quark fragmentation via a strange baryon (Y) is calculated as:
$P_{\lam}^q(Y) = - C_q^{\lam}(Y) P_q$,
where $C_q^{\lam}(Y)$ is the corresponding spin transfer coefficient, $P_q$
is the struck quark polarization which depends on the process. We use SU(6)
and BJ models to compute $C_q^{\lam}(Y)$.\\[-1cm] 

\subsection{Fragmentation model\label{sec:frag_model}}

\begin{wrapfigure}{l}{0.5\linewidth}
\vspace*{-1cm}
\begin{center}
\begin{tabular}{cc}
\epsfig{file=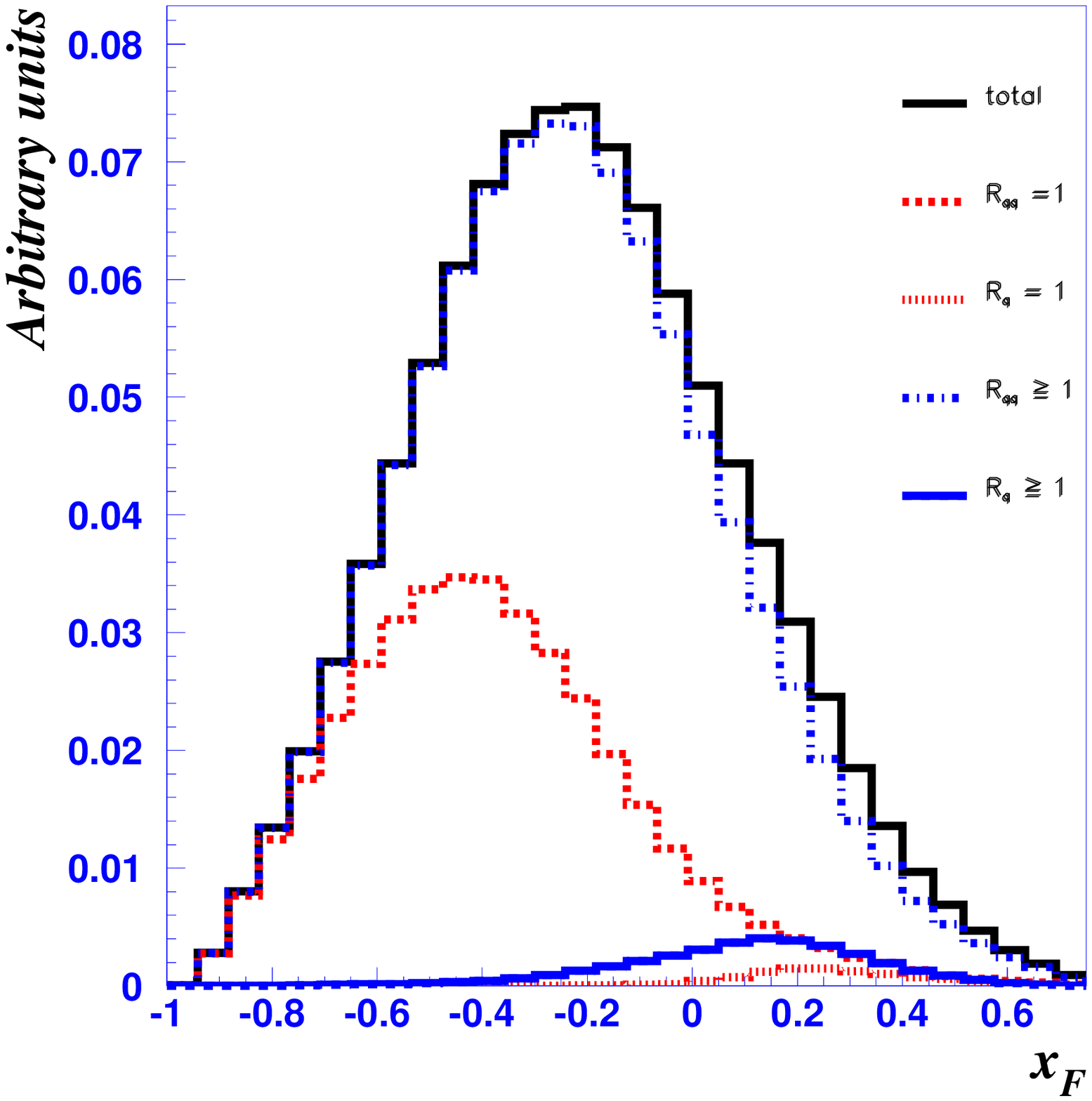,width=0.45\linewidth}&
\epsfig{file=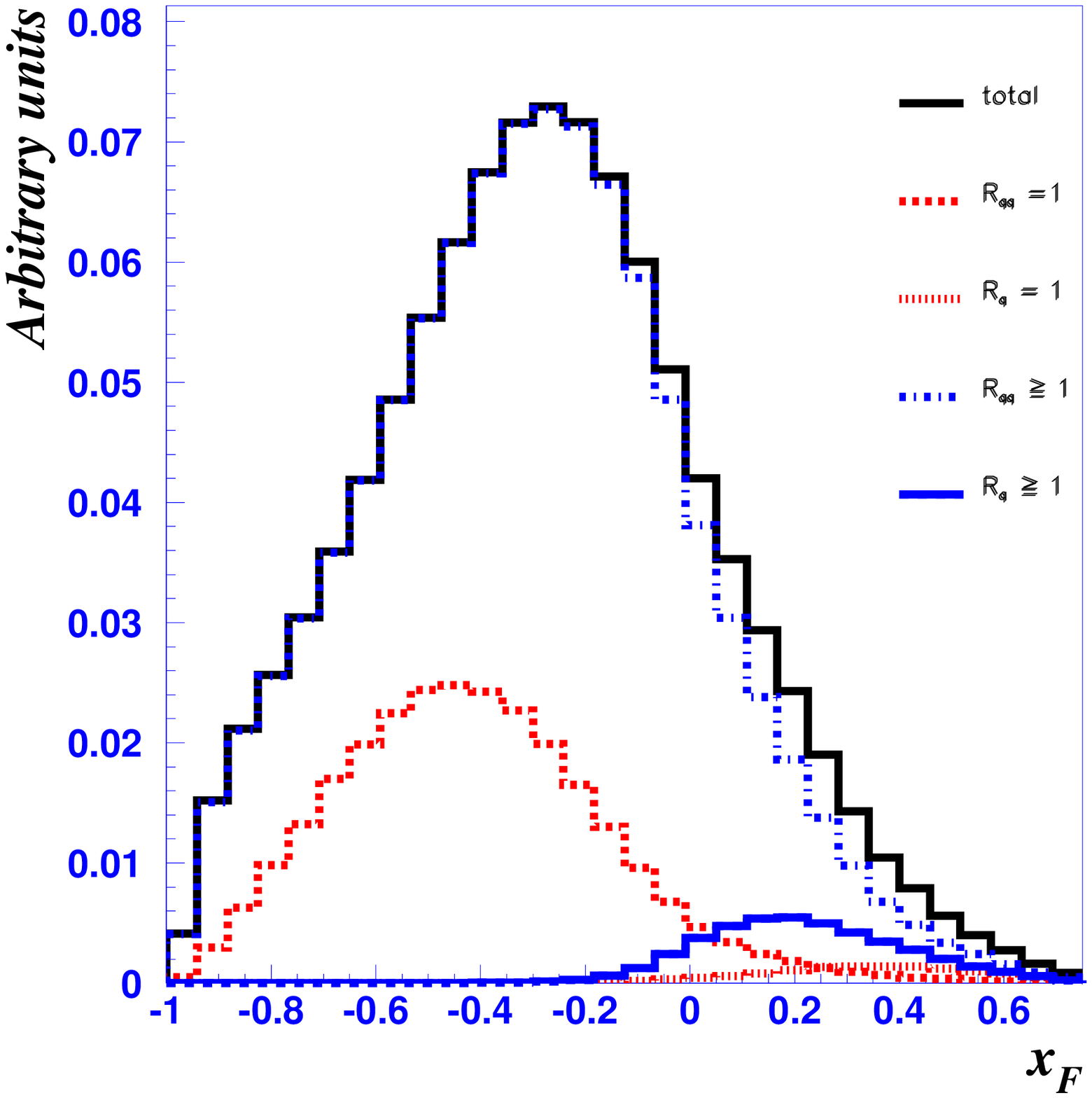,width=0.45\linewidth}\\
\end{tabular}
\caption{\label{fig:xf_default} 
Predictions for the
$x_F$ distributions of all $\lam$ hyperons (solid line), 
of those originating from diquark fragmentation and 
of those originating from quark fragmentation, for the two model variants 
A and B, as explained in the legend on the plots.
The left panel is for $\nu_\mu$ CC DIS with $E_\nu = 43.8$ GeV, 
and the right panel for $\mu^+$ DIS with $E_\mu = 160$ GeV. 
}
\end{center}
\vspace*{-1cm}
\end{wrapfigure}
To describe $\lam$ production and polarization in the full $x_F$ interval,
we use the LUND string fragmentation model, as incorporated into the {\tt
JETSET7.4} program~\cite{JETSET}. We use the {\tt LEPTO6.5.1}~\cite{lepto}
Monte Carlo event generator to simulate charged-lepton and (anti)neutrino
DIS processes. 
We introduce two rank counters: $R_{qq}$ and $R_q$ which correspond to the
particle rank from the diquark and quark ends of the string,
correspondingly. A hadron with $R_{qq}=1$ or $R_q=1$ would contain the
diquark or the quark from one of the ends of the string.
However, one should perhaps not rely too heavily on the tagging specified
in the LUND model. Therefore, we consider the following two variant
fragmentation models:

{\bf Model A:}
The hyperon contains the stuck quark (the remnant diquark) 
only if $R_q=1$ ($R_{qq} = 1$).

{\bf Model B:}
The hyperon contains the stuck quark  (the remnant diquark) 
if $R_q\ge 1$ and $R_{qq} \ne 1$ ($R_{qq}\ge 1$ and 
$R_q\ne 1$).

Clearly, Model B weakens the Lund tagging criterion by averaging over the 
string, whilst retaining information on the end
of the string where the hadron originated.

In the framework of {\tt JETSET}, it is possible to trace the particles'
parentage. We use this information to check the origins of the strange
hyperons produced in different kinematic domains, especially at various
$x_F$. According to the {\tt LEPTO} and {\tt JETSET}
event generators, the $x_F$ distribution of the diquark to $\lam$
fragmentation is weighted towards large negative $x_F$. 
{\em 
However, its tail in the $x_F>0$ region overwhelms the quark to 
$\lam$ $x_F$ distribution at these beam energies.}
In Fig.~\ref{fig:xf_default}, we show the $x_F$
distributions of $\lam$ hyperons produced in diquark and quark
fragmentation, as well as the final $x_F$ distributions.  These
distributions are shown for $\nu_\mu$ CC DIS at the NOMAD mean 
neutrino energy $E_\nu = 43.8$~GeV, and for $\mu^+$ DIS
at the COMPASS muon beam energy $E_\mu = 160$~GeV. The relatively small
fraction of the $\lam$ hyperons produced by quark fragmentation in the
region $x_F > 0$ is related to the relatively small centre-of-mass
energies - about 3.6 GeV for HERMES, about 4.5 GeV for NOMAD, about 8.7
GeV for COMPASS, and about 15 GeV for the E665 experiment - which
correspond to low $W$.

We vary the two correlation coefficients $C_{sq_{val}}$ and $C_{sq_{sea}}$
in fitting Models A and B to the following 4 NOMAD points:

1) $\nu p$: $P_x^\Lambda = -0.26 \pm 0.05(stat)$,
2) $\nu n$: $P_x^\Lambda = -0.09 \pm 0.04(stat)$,
3) $W^2<15$ GeV$^2$: $P_x^\Lambda(W^2<15) = -0.34 \pm 0.06(stat)$,
4) $W^2>15$ GeV$^2$: $P_x^\Lambda(W^2>15) = -0.06 \pm 0.04(stat)$.
We find from these fits similar values for both the $SU(6)$ and BJ models:
$C_{sq_{val}} = -0.35 \pm 0.05$, $C_{sq_{sea}} = -0.95 \pm 0.05$ (model A)
and $C_{sq_{val}} = -0.25 \pm 0.05$, $C_{sq_{sea}} = 0.15 \pm 0.05$ 
(model B).\\[-0.5cm] 

\section{Results\label{sec:results}}
\begin{wrapfigure}{r}{0.5\linewidth}
\vspace*{-1.5cm}
\begin{center}
\epsfig{file=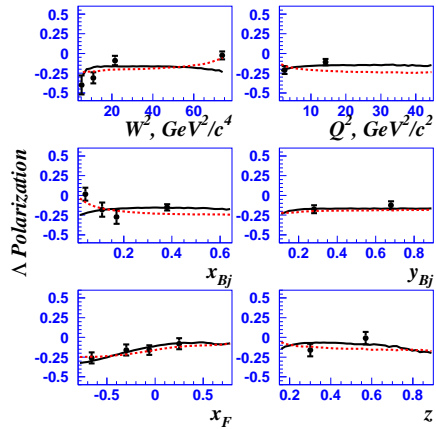,width=\linewidth}
\caption{\label{fig:numuxcc_6} The predictions of model A - solid line 
and model B - dashed line, for the polarization
of $\Lambda$ hyperons produced in $\nu_\mu$ charged-current DIS
interactions off nuclei as functions of $W^2$, $Q^2$, $x_{Bj}$,
$y_{Bj}$, $x_F$ and $z$ (at $x_F>0$). The points with error bars are
from ~\cite{NOMAD_lambda_polar}.} 
\end{center}
\vspace*{-0.5cm}
\end{wrapfigure}
\begin{figure}[htb]
\vspace*{-0.5cm}
\begin{center}
\begin{minipage}{0.45\linewidth}
\epsfig{file=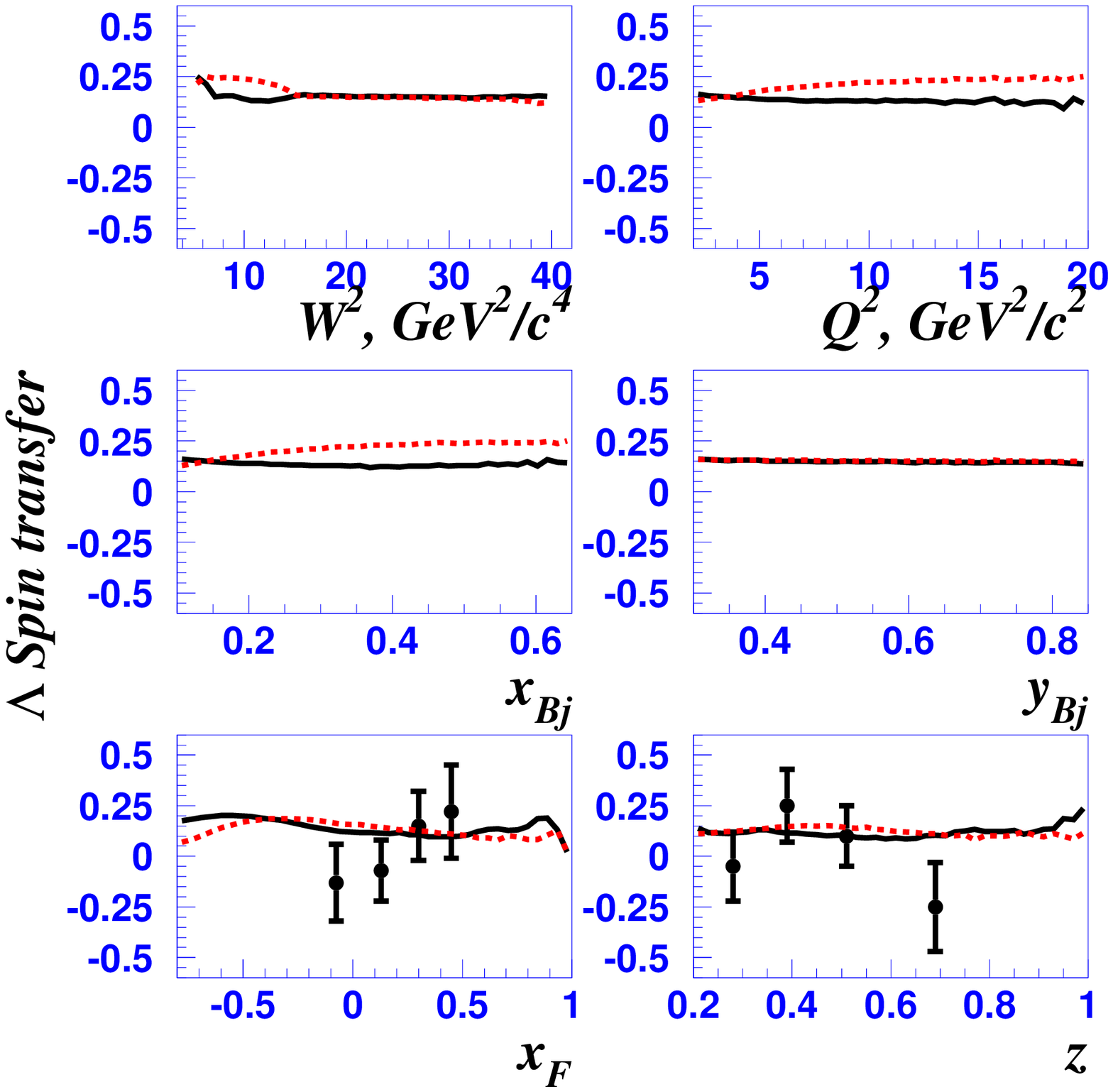,width=\linewidth}
\caption{\label{fig:e+xxxem_6} The predictions of model A - solid line, 
model B - dashed line, for the spin transfer to $\Lambda$ hyperons
produced in  $e^+$ DIS interactions off nuclei
as functions of $W^2$, $Q^2$, $x_{Bj}$, $y_{Bj}$, $x_F$ and $z$ (at
$x_F>0$). We assume $E_e = 27.5$~GeV, and the points with error bars are 
from~\cite{hermes}.} 
\end{minipage}
\hfill
\begin{minipage}{0.45\linewidth}
\epsfig{file=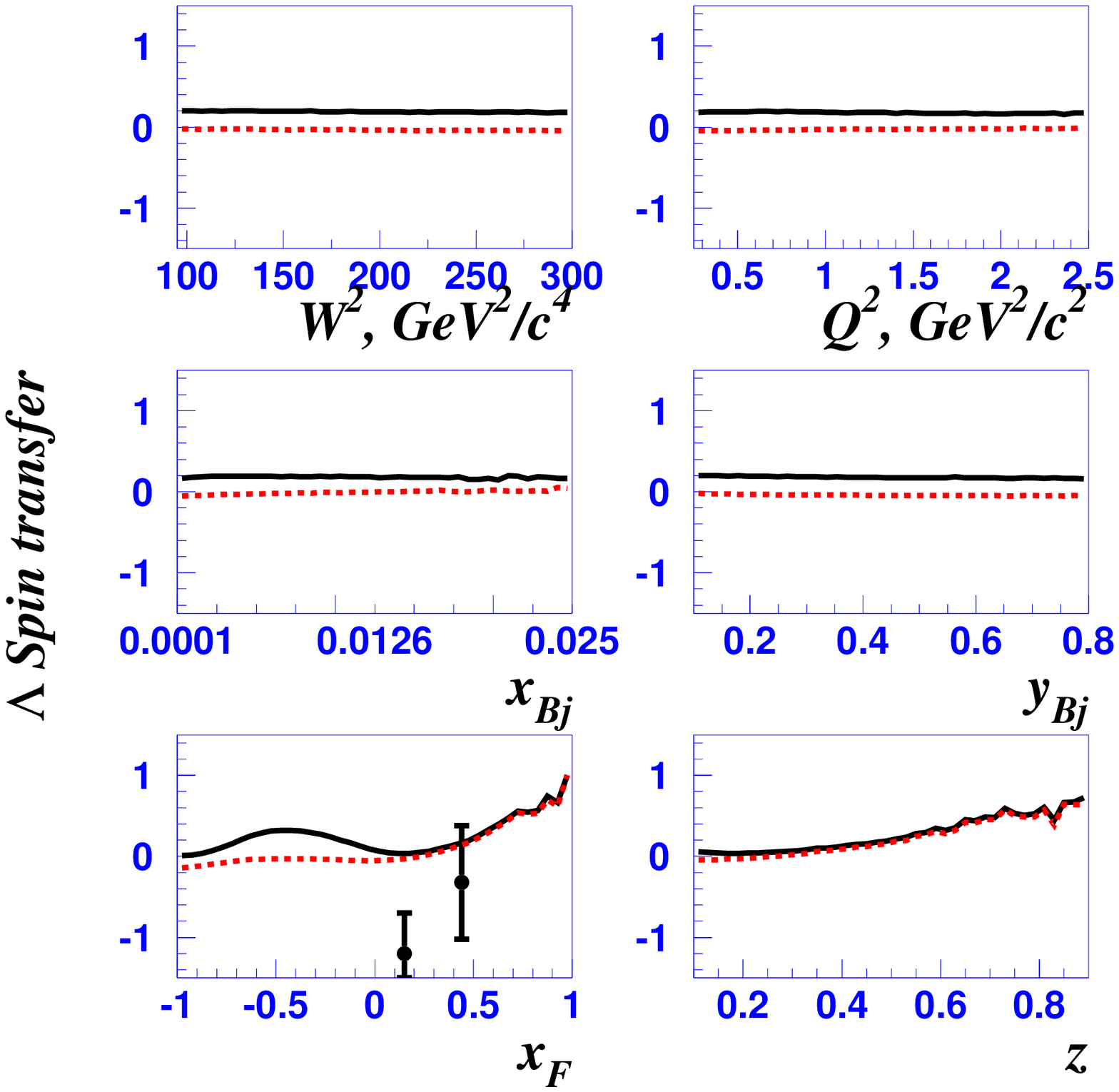,width=\linewidth}
\caption{\label{fig:e665_mu+xxem_6} The predictions of model A - solid 
line, model B - dashed line, for the spin transfer 
to $\Lambda$ hyperons produced in $\mu^+$ DIS
interactions off nuclei as functions of $W^2$, $Q^2$, $x_{Bj}$,
$y_{Bj}$, $x_F$ and $z$ (at $x_F>0$). Here we assume $E_\mu = 470$ GeV, 
as appropriate for E665~\cite{e665}.} 
\end{minipage}
\end{center}
\vspace*{-1cm}
\end{figure}

In Figs.~\ref{fig:numuxcc_6}, \ref{fig:e+xxxem_6}, \ref{fig:e665_mu+xxem_6}
we show our model predictions compared to the available data from the 
NOMAD~\cite{NOMAD_lambda_polar}, HERMES~\cite{hermes} and E665~\cite{e665}
experiments. One can conclude that our model quite well describes all 
the available experimental data.
The NOMAD Collaboration has measured separately the polarization of $\lam$
hyperons produced off proton and neutron targets.
We observe good agreement, within the statistical errors, between the 
model B description and the NOMAD data, whilst model A, although reproduces 
quite well the polarization of $\lam$ hyperons produced from an isoscalar
target, fails to describe target nucleon effects. We provide many possibilities
for further checks of our approach for future data 
(for details, see ~\cite{EKN}).

\end{document}